\documentclass[letter]{aa}  
\usepackage{graphicx}
\usepackage{txfonts}
\usepackage{natbib}
\bibpunct{(}{)}{;}{a}{}{,}

\newcommand{\HII}{{H\sc $\,$ii}}
\newcommand{\Cl}{[{Cl\sc $\,$ii}]}
\newcommand{\Fe}{[{Fe\sc $\,$iii}]}
\newcommand{\SIII}{[{\sc S$\,$iii}]}

\newcommand{\HA}{H\ensuremath{\alpha}}
\newcommand{\Ca}{{Ca\sc$\,$ii}}
\def\kms{$\mbox{km\,s}^{-1}$}

\def\farcs{\hbox{$.\!\!^{\prime\prime}$}}

\begin{document}
\title{Star-gas decoupling and a non-rotating stellar core in He~2-10}

   \subtitle{Integral field spectroscopy with
     FLAMES/ARGUS\thanks{Based on observations collected at the
       European Southern Observatory, Paranal, Chile, under observing
       programme 74.B-0771.}}

   \author{T. Marquart\inst{1}
          \and
          K. Fathi\inst{2,}\inst{3}
          \and
          G. {\"O}stlin\inst{3}
          \and
          N. Bergvall\inst{1}
          \and
          R. J. Cumming\inst{3}
          \and
          P. Amram\inst{4}
          }

   \offprints{T. Marquart}

   \institute{
   	     Department of Astronomy and Space Physics, Box 515, 
	     SE-75120 Uppsala, Sweden \\
              \email{thomas.marquart@astro.uu.se}
         \and
   	      Instituto de Astrof{\'\i} sica de Canarias, 
   	      C/ V{\'\i} a L{\'a}ctea s/n, 38200 La Laguna, Tenerife, Spain 
         \and
             Stockholm Observatory, AlbaNova University Center, 
	     SE-106 91 Stockholm, Sweden 
         \and
             Observatoire Astronomique Marseille-Provence, Laboratoire d'Astrophysique de Marseille UMR 6110, 2 place Le Verrier, F-13248 Marseille Cedex 4, France
	     }
   \date{Received; June 22, 2007}

  \abstract
   {}
   {We study the two-dimensional distribution and kinematics of the
     stellar and gaseous components in the centre of the prototype
     Wolf-Rayet blue compact dwarf galaxy, He~2-10. The aim is to
     compare the kinematics of gas and stars in order to determine
     whether they are consistent with one another, or if stars and gas
     can be decoupled due to gravitational perturbations and feedback
     from star formation.}
   {We have used the integral field unit ARGUS, part of FLAMES on the European
     Southern Observatory's Very Large Telescope, to target the \Ca\
     ${\lambda\lambda8498,8542,8662}$ \AA\ triplet in the central
     $300\times480$ parsecs of He~2-10. The selected wavelength regime
     includes several prominent spectral features, including the
     Paschen series and the \SIII\ emission-line, which we have used
     to derive the kinematics of the ionised interstellar medium.}
   {We find no systematic trend in the velocities of the stars over
     the observed field of view and conclude that the stellar
     kinematics is governed by random motions. This is in contrast to
     the motions the ionised interstellar medium, where we find spatial
     velocity variations up to 60 \kms. Our gas velocity field is
     consistent with previous studies of both the molecular gas and
     the feedback-driven outflow in He~2-10. We interpret the
     kinematic decoupling between the stars and the gas as He~2-10
     being in the process of transformation to a dwarf elliptical
     galaxy.}
   {}


   \maketitle

%________________________________________________________________
%________________________________________________________________

\section{Introduction}

Blue compact galaxies (BCGs) are important laboratories for testing
existing theories for dwarf galaxy formation and evolution. This is
because many of them form stars at rates that are unsustainable over
long periods and because of their low interstellar medium (ISM)
metallicities. The large amounts of young stars in BCGs give rise to a
blue continuum and strong emission-lines, dominating the optical
light. Their high star formation rates, high gas content, chaotic
morphology and low metallicities make them often being regarded as
local counterparts of young galaxies
\citep[e.g.,][]{2000A&ARv..10....1K}. However, detection of faint old
stellar populations suggests that these systems are intrinsically old,
even in the most youthful cases \citep{2005A&A...433..797O}.  The
puzzle of what triggers starbursts in BCGs remains unsolved, although
merger events are likely to play an important role.

Previously, Fabry-Perot interferometric studies of the \HA\ line
\citep{1999A&AS..137..419O,2001A&A...374..800O} have revealed
disturbed kinematics, indicating dwarf galaxy mergers as a likely
explanation for the formation of BCGs.  Despite local turbulence, the
\HA\ kinematics nevertheless suggest overall gravitational
support. Due to the ubiquitous presence of shocks, supernova winds and
heating by massive stars, it is unclear if the ISM follows the motions
of the stars, which are bound to follow the gravitational
potential. Both mass estimates from gas kinematics and evolutionary
scenarios based on the assumption that BCGs are self-gravitating may
therefore be uncertain.  This is of importance not only for the local
universe, but also at higher redshift, where strong emission-lines are
often the only usable kinematical tracers.  Thus, mass estimates of
high-redshift star-forming galaxies may change significantly if
the stars and gas display different kinematics.

Stellar motions are usually studied using optical absorption lines.
However, in BCGs, where metallicities are low and star formation is
strong, the absorption lines become easily diluted by the strong blue
continuum of hot stars and the nebular emission.  In the
near-infrared, the \Ca\ $\lambda\lambda8498,8542,8662$ \AA\ triplet
(hereafter Ca~triplet) can be used to derive reliable stellar
kinematic information \citep{1984ApJ...286...97D}. In a previous study
using Very Large Telescope (VLT) long-slit spectroscopy, we confirmed
on ESO\,400-G43 that the Ca~triplet can be used to probe the stellar
kinematics in BCGs \citep{2004A&A...419L..43O}. Our study showed that
the stellar component did not show the super-keplerian decline
observed for the ionised gas. Thus, stars can be kinematically
decoupled from the gas. With the help of integral field units (IFUs),
we can compare the gaseous and stellar motions by spectroscopy in two
spatial dimensions.

Here, we present results from IFU observations of the central
300$\times$480 parsecs in \object{He~2-10}, targeting the Ca~triplet.
He~2-10 is a well-known BCG with an absolute $B$-band magnitude of
$-19$ and two prominent star-forming regions (see
Fig~\ref{FigOverlay}). The super star clusters (SSCs) in the central
region (region A) have an age of 4-5 Myr \citep{2003ApJ...586..939C}
while those in region B, lying east of the centre, are older.  It has
been argued by \citet{1999A&A...349..801M} that strong feedback from
star formation drives a bipolar outflow in He~2-10. The systemic
velocity of 870 \kms\ is too low to serve as a good distance measure
because it cannot be assumed to belong to the Hubble flow, due to the
influence of the Virgo cluster. Here, we adopt the distance of 9 Mpc
used by \citet{1995AJ....110..116K}, corresponding to a physical scale
of 46.3 pc arcsec$^{-1}$.

\section{Observations and Data Reduction}
\label{sec:observations}

\begin{figure}
  \centering
  \includegraphics[width=.49\textwidth]{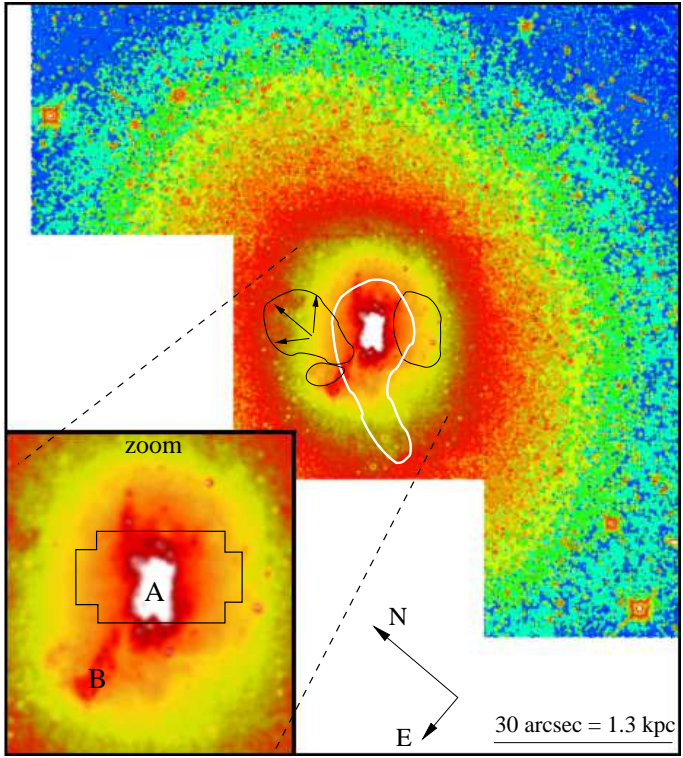}
  \caption{HST/WFPC2 F814W archival image of He~2-10 with an overlaid
    CO (thick white) contour from \citep{1995AJ....110..116K}. The
    thin black lines outline the positions of bubbles observed in
    H$\alpha$ emission \citep{1999A&A...349..801M}. The inset shows
    the central region at larger scale together with the footprint of
    our ARGUS observations and regions A and B, as discussed in the
    text.}
  \label{FigOverlay}
\end{figure}

The observations were carried out under good weather conditions during two
nights in November 2004 with FLAMES at European Southern Observatory's 
VLT (VLT/UT2) and its IFU, ARGUS, connected to the 
spectrograph GIRAFFE \citep{2002Msngr.110....1P}. 
The coarse (1:1) scale of ARGUS gives a spatial resolution of
$0\farcs52$ pix $\approx$ 30 pc/pix and the $22\times14$ pixels result
in a field-of-view of $11\farcs4 \times 7\farcs3$ placed with
$40^\circ$ position angle on region A in He~2-10 (see
Fig~\ref{FigOverlay}).  The light from each resolution element is
fibre-fed to the spectrograph where spectra with $R\approx 10400$
ranging from $8200$ \AA\ to $9380$ \AA\ were measured. The spectra
therefore cover the Ca~triplet, O~{\sc i} ${\lambda8446}$, \Cl\
${\lambda8579}$, \Fe\ ${\lambda8617}$, \SIII\
${\lambda9069}$\footnote{The rest-wavelength of this \SIII\ line is
  only determined to 0.5 {\AA} \citep{2007NIST}, corresponding to 16.5
  \kms. Throughout this paper, we adopt 9069.0 {\AA}.}, and the
Paschen 9 to Paschen 19 emission-lines of H~{\sc i}.

To save overhead time, we used fast target acquisition, thereby
trusting the pointing accuracy of the telescope. Four frames were
obtained with a total exposure time of 3.2 hours. The data were reduced
using the software and methods by \citet{2002A&A...385.1095P}, with
reduction steps including standard techniques for bias subtraction,
flat-fielding, wavelength calibration and optimised extraction of each
spectrum. The background emission from the sky was then subtracted
using the simultaneously observed spectra from the 15 sky fibres
located at a large distance around the target 
\citep[see][for details]{2003Msngr.113...15K}.

\begin{figure}
  \centering
  \includegraphics[width=.49\textwidth]{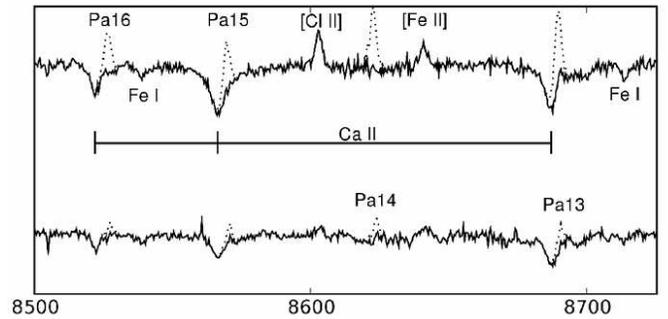}
  \caption{Example of two spectra from single fibres with
    different S/N. Dotted line: original spectrum. Solid line:
    after subtraction of the Paschen lines.
  }
  \label{FigSpectra}
\end{figure}

Since all three lines of the Ca~triplet have Paschen emission lines
superimposed on their red wings, it is crucial to accurately subtract
the emission features before deriving the stellar kinematics.  By
fitting several isolated Paschen lines simultaneously and using the
theoretical relative line-strengths, we subtracted a model spectrum for
Paschen emission for each pixel.  The model spectrum was constructed by
fitting the uncontaminated lines of Paschen 9, 10, 14, 17, 18, and 19.
Manual adjustments were made in cases where the signal was too low to
get a reliable model for the Paschen lines.  Fig.~\ref{FigSpectra}
shows two examples of subtracted and unsubtracted spectra in the
region around the Ca~triplet.  Both emission and absorption lines are
indicated.

We parametrise the stellar line-of-sight velocity distribution by a
Gaussian profile using the penalized pixel fitting algorithm developed
by \citet{2004PASP..116..138C} together with a selection of template
stars from the stellar library by \citet{2001MNRAS.326..959C}. Using
this technique, we derived the kinematics using the best linear
combination of 18 carefully chosen template stars covering M6~V with
$T_{\mathrm{eff}}$=3721 to O9~V with $T_{\mathrm{eff}}$=36300 K, and
metallicities ranging between -2.25 and 0.13 solar.
The stellar types used were
{K0~V}, 
{A8~Vn}, 
{K0~III}, 
{K3~IIb}, 
{K4~Iab}, 
{M3~II-III}, 
{K3~Iab}, 
{09~V},  
{B5}, 
{K3~III}, 
{A4}, 
{K0~V}, 
{M6~V}, 
{M2~V}, 
{G5~IIIw}, 
{K7~V}, 
{F0} and
{M4~V}.
Thus our derived kinematics is not affected by template mismatching
effects, since a different combination of stellar template spectra is
compiled for each individual spectrum. To match the template stellar
library, we degraded the spectral resolution of our data from
$0.2\,\AA/$pix to $0.85\,\AA/$pix. This simultaneously increased the
signal in the fainter pixels, while still resolving the Ca~triplet
well enough to derive the kinematics accurately. However, for fitting
the emission lines, we used the full-resolution spectra.

Although the pixel fitting routine delivers formal errors, these could
be underestimated since this formalism suppresses the noise in the
observations. We used Monte Carlo simulations to calculate more
realistic error estimates. For each pixel, we derived the kinematics
for 250 realisations, i.e., Gaussian randomised error-spectra added to
the galaxy spectra. We found that the average error over all pixels
is 6 \kms\ for the velocities, and 9 \kms\ for the stellar velocity
dispersions. We estimate that the emission line kinematics can be
derived within the same level of uncertainty or better.

In order to test the robustness of the derived stellar kinematics, we
were able to reproduce the stellar velocity field independently by
instead cross-correlating with three template stars of
  different type (K0II, K0III, G8Iab) that were observed in the same
nights with the same instrumental setup.  To quantify the systematic
error from the subtraction of the Paschen emission, we varied the
strength of the subtraction from twice the adopted amount to no
subtraction at all.  The goodness of the subsequent fitting confirmed
that the subtraction is optimal within 20\%. This corresponds to a
possible systematic error of 10 and 15 \kms\ for the stellar
velocities and line widths, respectively.  However, this is
  probably quite conservative since there is no indication that this
error acts always towards under- or oversubtraction and the spread in
stellar velocity values over the whole field is only 8 \kms.

\begin{figure}
  \centering
  \includegraphics[width=.49\textwidth]{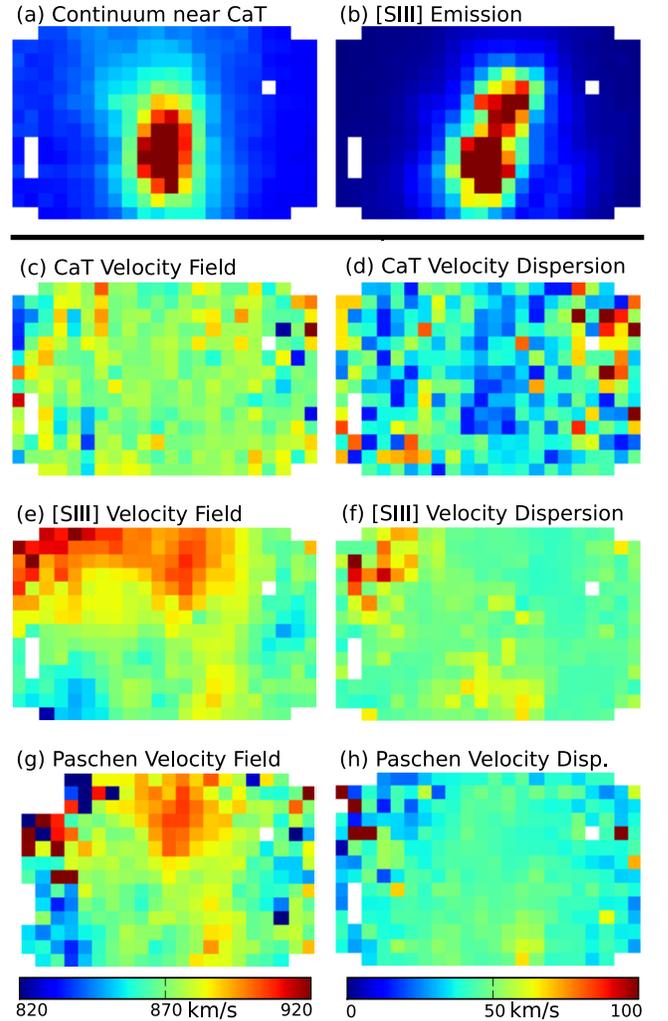}
  \caption{Panels showing the field of view of ARGUS, 22 by 14 pixels,
    corresponding to $11\farcs4 \times 7\farcs3$. As indicated in
    Fig.~\ref{FigOverlay}, left is north-east and up is north-west.
    Panels (a) and (b) show the continuum emission near the Ca~triplet
    and the monochromatic emission from \SIII\ respectively, in
    arbitrary flux units. Panels (c) and (d): Heliocentric
    line-of-sight velocity and velocity dispersion from the
    Ca~triplet. Colours for these and the following panels indicate
    values according to the colour bars beloweach column. Panels (e)
    and (f): Velocity and velocity dispersion from \SIII.  Panels (g)
    and (h): Velocity and velocity dispersion from the Paschen lines.}
  \label{FigPanels}
\end{figure}

\section{Results and Discussion}

\subsection{Stars}
\label{sec:stars}

The Ca~triplet is one of the most prominent near-infrared features of
cool stars. Although it originates mainly from red giant and
supergiant stars, it is present in stars of types A to M. Because the
light from the galaxy is dominated by the bright and young SSCs
\citep{2000AJ....120.1273J}, it is not the presumably relaxed old
underlying stellar population that is sampled when observing the
Ca~triplet, but instead fairly young stars --- red supergiants appear
after 5 Myr --- that have been formed during the current burst of star
formation. This is consistent with the fact that we do not see any
evidence for absorption in the Paschen lines (cf.~Fig.~\ref{FigSpectra}).

The stellar velocity field (panel (c) in Fig.~\ref{FigPanels})
displays no significant velocity gradient across the field, and
thereby no sign of rotation.  We find the stellar component of He~2-10
to have a systemic velocity of $872\pm 6$ \kms, in good agreement with
the usually adopted systemic velocity for He~2-10
\citep[e.g.,][]{1987A&A...182..179J}.  The absence of a gradient in
the stellar velocity field suggests that the young component, which
dominates the light, is relaxed and governed by random motions. The
stellar velocity dispersion map (panel (d)) shows an average value of
$45 \pm 4$ \kms\ outside the bright starburst centre (region A), while
in the very centre itself the stellar velocity dispersion drops to $28
\pm 3$ \kms.  This can be understood by assuming that the young stars,
dominating the light, are being formed in a locally dense and
dynamically less hot environment. Outside the burst region, the stars
are older and have had more time to virialize with the starburst host,
thereby increasing the velocity dispersion.

Assuming virialisation, we can estimate masses using the
  relation $M_\sigma = 1.1 \times
10^{6}\,\cdot\,r_{e}\,\cdot\,\sigma^2$, where the effective radius
$r_{e}$ is given in units of kpc, $\sigma$ in \kms, and $M_\sigma$ in
$M_\odot$ \citep{1992ApJ...399..462B,1996ApJ...460L...5G}. For the
central starburst with a size of $\approx 1.1\arcsec$ (derived from
archival Hubble Space Telescope NICMOS camera $H$-band photometry) and
$\sigma = 28$ \kms, we calculate the mass to be $4\times
10^{7}M_\odot$.  The same exercise for the region outside region A,
represented by an effective radius of $\approx 6 \arcsec$ and the
average stellar velocity dispersion $\sigma=45$ \kms\, gives a total
galaxy mass of $6\times 10^{8}M_\odot$. This compares very well with
the total mass in atomic and molecular gas measured by
\citet{1995AJ....110..116K}. The irregular structure of the central
morphology and possible variations in the mass to light ratio leads to
an uncertainty in $r_e$ of about a factor of 2, translating linearly
to the derived mass. By integrating the light profiles from
\citet{2003A&A...410..481N}, we estimate a mass-to-light ratio in the
$H$ band of $0.05$ and $0.2$ for the two masses derived above,
respectively. This is consistent with the very young stellar
population in He~2-10.

\subsection{Gas}
\label{sec:gas}
The emission-line velocity dispersions of the \SIII\ and Paschen
emission-lines are almost constant over the field of view, with
average values $49 \pm 6$ (standard deviation) and $44 \pm 3$ \kms,
respectively.  In the northern corner of the field, there is an
indication of an increase in $\sigma$ for the \SIII\ and a decrease
for the Paschen lines, although this is the region with the weakest
signal. Overall, these values are higher than the stellar velocity
dispersion at the very centre, but they agree with the value outside
the starburst region. The observed gaseous velocity dispersion is more
than a factor two higher than the average measured turbulent velocity
dispersion of \HII\ regions in spiral galaxies
\citep[e.g.,][]{1994ApJ...425..720C} but not unusually high for a BCG
where it is often used empirically as a mass measure
\citep[e.g.,][]{1987MNRAS.226..849M,1996ApJ...460L...5G,2001A&A...374..800O}. 
Since the gas velocity dispersion in He~2-10 is higher than that of
the stars in the centre, the outflow (see below) probably also
contributes to the line width.

The velocity fields of the \SIII\ and Paschen emission-lines (panels
(e) and (g) in Fig.~\ref{FigPanels}) clearly deviate from that of the
stars.  We observe gas velocities to be generally higher than the
stellar velocity and find the region west and north-west of the
central starburst to display velocities reaching up to 50 \kms\ higher
than the systemic velocity. At the east and south-east, the \SIII\
line-of-sight velocity is less than the systemic value by 10 \kms.
Except in regions with barely detected Paschen emission, the velocity
field of the \SIII\ line resembles that of the Paschen lines, 
in agreement with H$\alpha$ velocities measured by
\citet{1999A&A...349..801M} in the sense that we observe blueshifted
lines towards the north-east of the field. We also find agreement with
our H$\alpha$ Fabry-Perot measurements \citep[][and 2007 in
preparation]{2006IAUS..235E.267M}.

Although on larger scales, the gas in He~2-10 is known to rotate
\citep{1995AJ....110..116K,2006IAUS..235E.267M}, the emission-line
velocity fields we observe here alone cannot establish whether the
high and low values to either side of the zero-velocity curve are
signatures of rotation in the centre of He~2-10. The rough orientation
of the velocity gradient with the morphological major axis might
suggest this, but the corresponding mass from assuming Keplerian
motions is an order of magnitude below the value derived above. In
addition, the lack of rotation in the stellar component seems to
preclude the existence of a dominant star-forming disk.
\citet{1995AJ....110..116K} find an elongated cloud of molecular gas
with a tail towards the south-east (see also our
Fig.~\ref{FigOverlay}). Over the region of our field of view, the CO
gas displays a velocity gradient that is in agreement with the gaseous
motions presented here. Since the stars do not follow these motions,
they probably have not formed out of the same gas, which might instead
represent the part of the cloud with high angular momentum or the
infall of fresh gas.

On the other hand, the ISM is subject to strong feedback from the
starburst that drives the well-studied outflow in He~2-10. Both
\citet{1999AJ....117..190B} and \citet{1999A&A...349..801M} have
observed bubble structures in deep H$\alpha$ images (cf.\ Fig.\
\ref{FigOverlay}) blowing out towards the north-east and south-west of
the central starburst.  \citet{2000AJ....120.1273J} find this scenario
to be consistent with the measured energy output from the SSCs in the
central starburst. We measure blueshifted lines towards the proposed
outflow regions on both sides that can be attributed to a bipolar
outflow, seen close to edge-on.

\citet{2006ApJ...646..858S} have used ultraviolet (UV) absorption
lines to measure the speed of the outflow at the position of the
brightest SSCs and found it to be $-$170 \kms\ with respect to the
systemic velocity. In contrast to this, we measure gaseous velocities
at the systemic value. We attribute this discrepancy to the different
regions probed. UV absorption lines sample the cold and warm ISM in
the foreground, while emission lines trace the ionised ISM at the
location of the starbrust.

\section{Conclusions}
\label{sec:decoup}

We have used integral field spectroscopy to measure the stellar
kinematics in He~2-10 using the Ca~triplet absorption lines, and
compared it with the motions of the ionised gas. We find a large
discrepancy between stars and gas. The stellar component shows no
signatures of rotation or any other non-random motions.  The gas,
however, shows bulk motions that are consistent with previous results
from molecular gas kinematics and the bipolar outflow, driven by the
starburst in the centre.

In particular, IFU spectroscopy has allowed us to show that the stars
in the centre show no sign of rotation along any direction and that
gas and stars are kinematically clearly decoupled.  The virialization
of the stellar component is compatible with He~2-10 being an advanced
merger because simulations show that virialization is quick and that
the peak in star formation rate in a galaxy merging event coincides
with the completion of the merger
\citep[e.g.,][]{1992ARA&A..30..705B,2006MNRAS.373.1013C}.

That the stellar component is supported by velocity dispersion rather
than rotation also hints toward the future evolution of He~2-10 in the
direction of an elliptical-like galaxy \citep{2002AJ....124.3073G}, as
earlier suggested by \citet{1995ApJ...440L..49K} for more distant
emission line galaxies.  Indeed, the strongly peaked light profile
will probably, after the burst has faded, resemble that of a nucleated
dE like M~32 or NGC~205. The lower velocity dispersion that we find
for the youngest stellar component in the very centre of He~2-10 may
in time be dynamically heated and begin to resemble the rest of the
galaxy.

Our results also suggest a need for caution in studies of starburst
galaxies at higher redshifts, where emission-lines often are the only
feasible kinematical tracers. Observed spatial variations in
line-of-sight velocities need not be representative of the
gravitational potential if decoupling between gas and stars is
common.

\begin{acknowledgements}
  T.~Marquart wishes to thank Nikolai Piskunov for assisting in the
  data reduction. K.Fathi acknowledges support from the Instituto
    de Astrof\'isica de Canarias through project P3/86, the Royal
    Swedish Academy of Sciences' Hierta-Retzius foundation, and the
    Wenner-Gren Foundation. N.~Bergvall and G.~\"{O}stlin acknowledge
  support from the Swedish Research Council.
\end{acknowledgements}

\bibliographystyle{bibtex/aa}
\bibliography{argus}

\end{document}